# Spam Detection Using BERT

Thaer Sahmoud[1], Dr. Mohammad Mikki[2]
[1]Thaer.sahmoud@students.iugaza.edu.ps, [2]Mmikki@iugaza.edu.ps
[1,2]Computer Engineering Department, Islamic University of Gaza, Palestine

*Abstract- Emails and SMSs are the most popular tools in today communications, and as the increase of emails and SMSs users are increase, the number of spams is also increases. Spam is any kind of unwanted, unsolicited digital communication that gets sent out in bulk, spam emails and SMSs are causing major resource wastage by unnecessarily flooding the network links. Although most spam mail originate with advertisers looking to push their products, some are much more malicious in their intent like phishing emails that aims to trick victims into giving up sensitive information like website logins or credit card information this type of cybercrime is known as phishing. To countermeasure spams, many researches and efforts are done to build spam detectors that are able to filter out messages and emails as spam or ham. In this research we build a spam detector using BERT pre-trained model that classifies emails and messages by understanding to their context, and we trained our spam detector model using multiple corpuses like SMS collection corpus, Enron corpus, SpamAssassin corpus, Ling-Spam corpus and SMS spam collection corpus, our spam detector performance was 98.62%, 97.83%, 99.13% and 99.28% respectively.*

## 1. INTRODUCTION

Email is the most important tool for communications and it's widely used in almost every field like business, corporations, education institutes and even for individual users. To communicate effectively spam detection is one of the important features that aim to enhance user experience and security. Spam mail, or junk mail, is a type of email that is sent to a massive number of users at one time, frequently containing cryptic messages, scams, or most dangerously, phishing content. A spam detector -as shown in figure 1-is a program used to detect unsolicited, unwanted and virus-infected emails and prevent those messages from getting to a user's inbox.

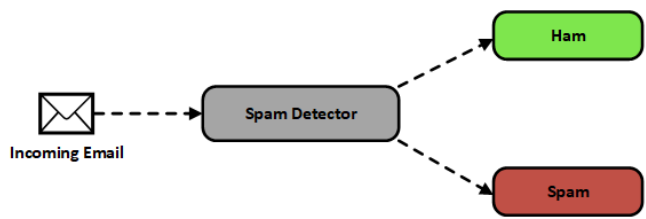

*Figure 1: Spam Detector*

Spam detectors use various methodologies for email classification, one methodology is to filter email based on the sender address which known as Blocklist filtering where a list of emails for spammers and that list is updated frequently to keep up with spammers who change their email address, however when spammer change his email domain that methodology will no work until the new spammer email is listed in the blocklist, so we cannot rely on that method only.

The other methodology is for spam detector is to filtering according to email content where spam detector checks the content of each email and classify it, and that type of spam detectors has very high performance since spam emails content is often predictable, offering deals, promoting explicit content or targeting basic human feelings, such as desire and fear. And another methodology is to have a rule-based email filtering that allow the user to manually configure a special rule for all incoming emails and classify it as spam or not, an example of these rules is to classify emails from specific sender of the email subject contains a certain phrase or even the message contains some words and whenever an incoming email matches one of the rules, it automatically forwards the email to a spam folder.

There are two approaches to implement spam detector, machine learning based approach and non-machine learning based approach, for non-machine learning base approach, spam detection was to manually construct document classifiers with rules compiled by domain experts. The problem with non-statistical approaches is that there is no learning component to admit messages whose content 'look' legitimate and that leads to undetected spam emails and therefore the detector accuracy will be low. Due to the above limitations, many researchers have used machine learning algorithms to build their spam detectors, so by trying the spam detector model on a train dataset it can predict spam emails for emails that have the same pattern as in that dataset. The most common spam emails are advertising messages that makes up 36% of the all-world spam content **[1]**. Sometimes advertising spams presence can be interpreted as disrespectful and annoying, however, as people generally don't like to see their email databases swarmed by hundreds of email messages advertising elixirs of youth and restaurant chains. Adult-related content is the second-biggest spam category, accounting for roughly 31.7% of all spam messages, 26.5% of all unwanted emails are related to financial matters, the third-biggest spam email category. And the most dangerous of spams is Scams and fraud account which is about 2.5% of all spam emails, however, phishing statistics indicate that identity theft is the goal of 73% of those, and for every 12,500,000 emails sent, spammers receive one reply [2].

In this research we use Natural Language Processing "NLP" pre-trained model named Bidirectional Encoder Representations from Transformers "BERT" to build a high-performance spam detector by training the model on different datasets. BERT [3] is a machine learning pre-trained model that is used for Natural Langue Understanding task to help machines understand the context of sentences which was developed by Google AI in 2018.

BERT use transfer learning and the architecture is based on transformer model [4], transfer learning means that train a model for a general task and then use that knowledge obtained from the general task training to fine tune BERT to a new task. BERT has been trained on two tasks: masked language task where sentences are fed to the model and some words are masked or hidden for the model and let the model try to predict these hidden words, and the other task is sentence prediction where pair of sentences are fed to the model each round and the model need to predict wither one sentence is followed by the other or not. BERT has been trained over a large dataset for these two tasks, the dataset contained all English Wikipedia and 11,038 books. BERT use encoder from transformer model that is a type of neural network which it takes a sentence as input to the model then each word of the sentence is then tokenized and these tokenized words are fed to the BERT model and BERT output is a vector representation for each tokenized word. Using encoders from Transformer enable BERT to have a better context understanding than traditional neural networks such as LSTM [5] or RNN [6] since encoder process all inputs which is actually the whole sentence simultaneously so when building a context for a word BERT will take in account the inputs before it and also the inputs after the word, while the LSTM or RNN process the input taking in account only the prior inputs, and that will be reflected on the output vector value for the word, so the word "apple" in the two sentences (I need the apple) and (I need the apple product) would have the same vector value -and as a result the same meaning- when using LSTM or RNN, on the other hand it would have two different vectors using BERT, so as a result, using BERT will -in most cases- give us a better performance than using the traditional machine learning algorithms.

The rest of paper is designed as follow, in section 2 we will define the problem of spam emails and the financial impact for spams, section 3 we will discuss the related works in spam detection field and what methodology did other researchers used in that field, in section 4 we will in details discuss our model and how we fine-tuned BERT for our spam detection task, our model results and performance for each corpus that we use to train and evaluate our model is listed in section 5, and finally the paper conclusion is in section 6.

## 2. PROBLEM STATEMENT

Spam is a large issue for almost all internet users, since according to statistics [7] spam 52% of the participant polled in the survey stated that spam was a major problem. According to Kaspersky labs [8] On average, 45.56% of global mail traffic was spam in 2021.

Choudhari [9] stated that on average about 122.3 billion spam emails were sent daily, and according to [10], FBI reported a loss of 12.5 billion USD to business email customers, the FBI's Internet Client Complaint Center "IC3" [11] stated that the total global loss exceeded 43 billion USD for interval between June 2016 and December 2021, and according to Verizon's 2019 Data Breach Investigations Report (DBIR) [12] which include an investigation of malware concluded that a whopping 94% used email as a delivery method.

All these facts about spam emails motivate us to implement an effective spam detection model that use BERT which is an NLP model to enhance our spam detector performance.

## 3. RELATED WORKS

Many research and efforts discussed spam emails and how implement an effective spam detector using machine learning algorithms, Harisinghaney [13] tried to implement text and image-based spam emails with the help of the k-nearest neighbor (KNN) algorithm, Naive Bayes, and reverse DBSCAN algorithms, Laorden [14] developed a Word Sense Disambiguation preprocessing step before applying machine learning algorithms to detect spam data. Finally, results indicate a 2 to 6% increase in the precision score when applied on Ling Spam and TREC datasets. Janez-Martin [15] made the combined model of TF-IDF and SVM showed 95.39% F1score and the fastest spam classification achieved with the help of the TF-IDF and NB approach. Faris [4] proposed a Particular swarm optimization PSO-based Wrapper with a Random Forest algorithm that effectively detects spam messages. Marie-Saint [16] introduced Deep Learning-based approaches Deep learning mimics the human brain to solve the given the firefly algorithm with SVM and worked with Arabic text. This article concluded the proposed method outperformed SVM alone.

[17] built a spam detection model for Chinese, their model based on long-short attention mechanism that convert words to vectors based on the context of the sentence, they trained the model using Trec06C which is a Chinese spam dataset, their model achieved high performance with accuracy of 99.3, [18] use modified Transformer model for SMS spam detection, they modified Transformer by adding memory and a linear layer with a final activation function that takes the output of Transformer model as an input to the final classification layer, they trained their model using two datasets, the first dataset is SMS Spam Collection v.1 [19] and the other dataset is UtkMl's Twitter Spam Detection Competition [20] from Kaggle, the model f1-score was 98.92 for SMS dataset and 94.51 for twitter dataset [21] use fine tune BERT for spam detection task, they used various datasets to train and evaluate the model, model performance

for each dataset was as follows: SpamAssassin dataset with f1-score of 97, Enron dataset achieved f1-score of 97, Ling-Spam dataset f1-score was 94 and SpamText dataset f1-score was 96.

## 4. PROPOSED MODEL

There are mainly two types of BERT which are BERT base and BERT large, the BERT base consists of a stack of 12 encoders from Transformer model above each other's and a hidden size of 768 as shown in figure 2, while BERT large consists of stack of 24 encoders and the hidden size is 1024, we use BERT base model to build our Spam detector. Transformer encoders are all identical in structure and consists of a self-attention layer followed by a feed forward neural network layer, the self-attention layer is the main part of BERT – and also Transformer- is responsible for making the encoder while encoding a word to take in account the other words in the input sentence, so the word embeddings will be based on the sentence context, the output of the self-attention is then fed to a feed forward neural network layer to process the output from one attention layer in a way to better fit the input for the next attention layer.

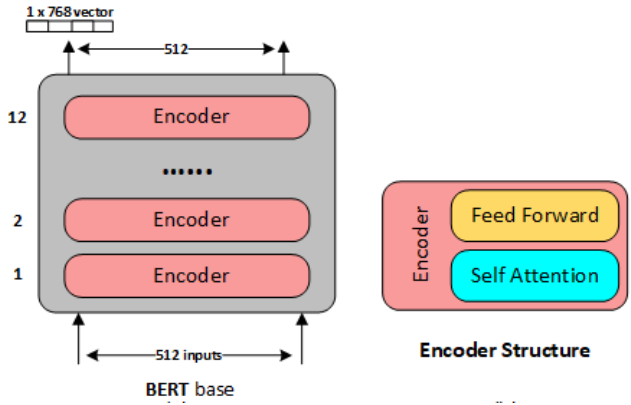

*Figure 2: BERT base architecture*

To build an effective spam detector we fine-tune BERT base by adding a classification fully connected layer above the BERT pre-trained model as shown in figure 3, we then take the output of the final encoder and fed it to a single fully connected layer to accomplish a binary classification task, this process is called fine-tunning the BERT model, the input of our classification layer is the hidden state and the output is the classification result whether it's spam or ham.

Our spam detector works as follows, first the email is tokenized using BERT tokenizer which is also called wordPiece tokenizer [22] that split the sentence into chunks of words, some words as split as a single word and some are split into multiple words base on the tokenizer vocabulary file, after tokenization the tokens are padding with a special token named "PAD" to max BERT input size which is by default 512 and it's configurable, after that two special tokens are added the "CLS" token which stands for classification is added at the beginning of the sentence and "SEP" token which stands for separation is added at the end of the sentence, the total number of tokens ("CLS", email tokens, padding tokens "PAD" and the "SEP" token) which are fed to BERT that are fed to BERT must equals the max BERT input size.

On the next step, the total input tokens are converted to integer IDs based on the tokenizer vocabulary file and these IDs are fed to the BERT model with a matrix of ones and zeros that indicate whether the token input is a real input so the corresponding element in that matrix will be set to one, or it's a padded input with corresponding element equals zero in that matrix, the BERT model is then convert input tokens to a vector of size equals BERT hidden size which is 768 in our case since we use BERT base, we then take that last hidden state for CLS token that represent the whole sentence classification and fed it to the classification layer that decide wither the input email was a spam email or ham email.

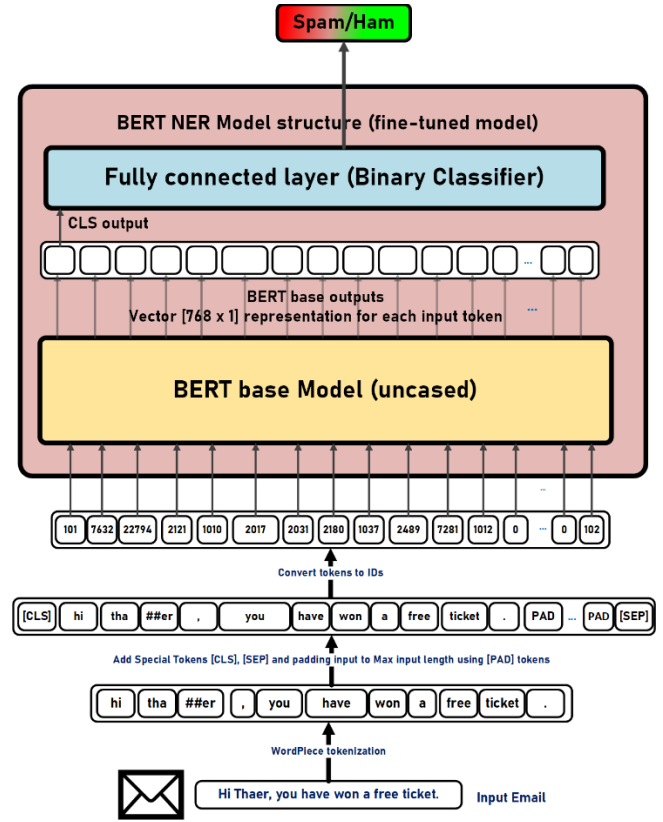

*Figure 3: Proposed spam detector model*

## 5. TRAIN THE MODEL

To train the model we use the following corpuses, SpamAssassin corpus [23], SMS Spam Collection v.1 [19], Enron corpus [24] and Ling-Spam corpus [25]. SpamAssassin is a public email corpus that's used to train and evaluate spam detectors, the corpus consists of 5,922 emails where each email is labeled as spam or ham, where 4,096 emails are ham and 1,853 emails spam. We also use SMS Spam Collection v.1 corpus that contains 5,572 sample SMS that are divided into 4,825 ham SMS and 747 spam SMS, beside these two corpuses we use Ling-Spam corpus that was created by [25], the Ling-Spam dataset corpus contains both legitimate and spam emails., in English language and the last corpus that we use is Enron corpus, this corpus was collected and prepared by the CALO Project (A Cognitive Assistant that Learns and Organizes), the corpus contains 9,972 emails where 4,999 emails are spam and 4,972 emails are ham, table 1 shown below illustrate details of corpuses that we use.

*Table 1: Corpuses details*

| Corpus | Emails/SMSs | Spam | Ham |
|---|---|---|---|
| SMS Spam Collection v.1 | 5,572 | 747 | 4,825 |
| SpamAssassin | 5,796 | 3,900 | 1,896 |
| Ling-Spam | 2,893 | 481 | 2,412 |
| Enron | 33,716 | 16,852 | 16,493 |

For each corpus of these three corpus, the first step is to clean up the data, so we remove URLs form emails and any unwanted characters and if the email body contains html tags we just remove these tags so we can focus on the text itself, after corpus pre-processing, we split each corpus into train dataset and evaluation dataset, train dataset is use to train the model on spam detection task or in other words to tune model parameters or weights -BERT base has around 110M weights-, model training is done as follow emails and the corresponding labels are fed to the model, we use PyTorch which is an open source machine learning framework to train the model which enables us to process multiple emails on parallel, so we divided the train dataset into batches with batch size of 16 emails and the corresponding labels, at the start of training, the model has initial weights, the training process is divided into two sub-processes, the first sup-process is a feed-forward process and backpropagation sub-process, in feed-forward, the input email is processed with weights that are distributed between the model layers and in the final stage of feed-forward the model produce an output related to that input email, we then compute the difference between the model output and the output that is in the corresponding input label and this difference is then used to tune the model weights through a backpropagation started from tunning the output weights until reach the input weights, then the second input is fed to the model and the feed-forward and backpropagation process is repeated to tune the previously updated weights, to optimize weights update we use Adam optimizer which is responsible for choosing the optimal update value for each weight. We trained the model using 10 epochs that mean the whole dataset are used to train the model for 10 times.

The evaluation process, we feed the model with evaluation dataset that contains emails and the corresponding labels, but this time there is no weighs update process, emails are fed to the model and when model classifies the input email, we compare the model classification with the corresponding label for the input to compute model performance. The best way to calculate the model performance is to first compute the confusion matrix **[26]** which is illustrated in figure 4, the Gold labels are the origin labels of the dataset and system prediction labels are the output of our classification model, since we have to classify the model to two classes (ham/spam) then we calculate the confusion matrix and f1-score for each class, that means we firstly calculate for "Spam" class, in this case, true positive is that the email was spam and our model classify it as spam, while false positive is that the input label was ham but our model classifies it as a spam, false negative is that out system classifies it as a ham but the correct label is spam, while true negative that the model classify the email as ham email and it was labeled in the dataset as ham, secondly we calculate the confusion matrix and f1-score for the "Ham" class, and in that case true positive is the total number of inputs that the model classify it as ham and it's labeled as ham in the dataset, false positive is the total number of inputs that the model wrongly classify them spam but they are labeled as ham in the dataset, while false negative is the total number of inputs that the model wrongly classify them as ham but they are labeled as spam, and finally true negative is the total number of inputs that are labeled as spam and the model predict them as spam too.

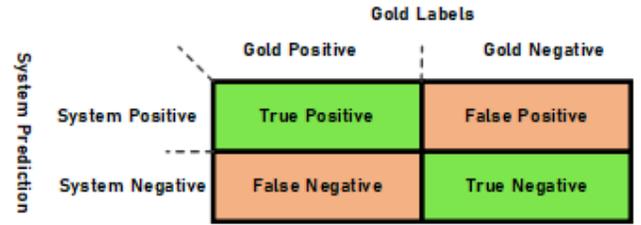

*Figure 4: Confusion matrix*

From these confusion matrices we calculate three measures for each class, the first measure is the Precision that measures the percentage of the items that the system detected that are in fact positive and is calculated as in equation 1.a, the second measure is the Recall that measures the percentage of items actually present in the input that were correctly identified by the system and calculated as in equation 1.b. From recall and precision, we calculate the f1-score comes from a weighted harmonic mean of precision and recall and it measures the model performance, equation 1.c is used to calculate f1-score.

$$\boldsymbol{Precision} = \frac{true\ positive}{true\ positive + false\ positive} \qquad Equ.\ 1.a$$

$$\boldsymbol{Recall} = \frac{true\ positive}{true\ positive + false\ negative} \qquad Equ.\ 1.b$$

$$\boldsymbol{F1\ Score} = \frac{2*Precision*Recall}{Precision + Recall} \qquad Equ.\ 1.c$$

The total model performance values "precision, recall and f1-score" are the average values of the two classes.

## 6. RESULTS

After training out model using training dataset for each corpus, we evaluated the model using the evaluation dataset from the same corpus and the results was as follows

### 6.1. ENRON CORPUS RESULTS

We train our model on Enron corpus, we split that corpus into 90 % for training dataset and 10% for evaluation, figure 5 shows the confusion matrix and from that confusion matrix we can see that our spam detector model detects 812 emails correctly as ham out of 822 ham emails and also classifies

833 emails correctly as spams out of 846 spam emails, table 2 shows the "Recall", "Precision", and "F1-scores" for each class and the total model performance.

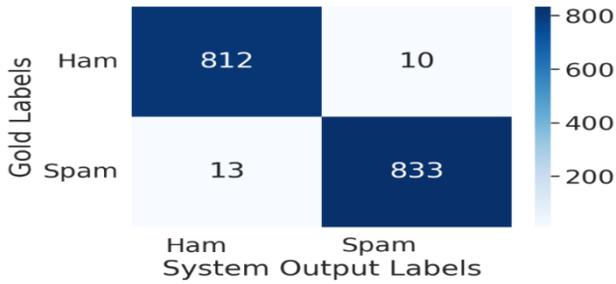

Figure 5: Confusion Matrix for Enron Corpus

The support column in table 2 indicates number of emails that belongs to the specific class in the evaluation dataset, for calculating the overall results we have two methods, the first method macro average which means we simply take the average for every class, and the other method is the weighted average where each performance value is weighted with the support column, and we use the weighted average to get the overall model performance.

Table 2: Model performance for Enron Corpus

| Class | Precision | Recall | F1-score | Support |
|---|---|---|---|---|
| Ham | 0.9842 | 0.9878 | 0.986 | 822 |
| Spam | 0.9881 | 0.9846 | 0.9864 | 846 |
| Weighted-average | 0.9862 | 0.9862 | 0.9862 | 1,668 |

As we can see in table 2, the model achieves overall f1-score of 98.62%.

### 6.2. SPAMASSASSIN CORPUS RESULTS

For SpamAssassin corpus, we also divide the corpus into train dataset that contains 90% of total corpus, and 10% to evaluate the mode performance, the resulted confusion matrix was as shown in figure 6, our model classifies 631 emails correctly out of 634 as ham emails, and classifies 277 emails as spam out of 294 spam emails in evaluation dataset, and it achieves f1-score of 97.83%

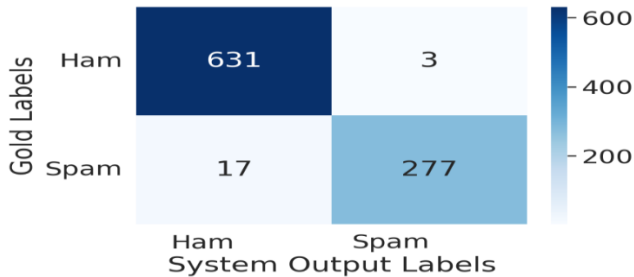

Figure 6: Confusion Matrix for SpamAssassin Corpus

### 6.3. LING-SPAM CORPUS RESULTS

We divide the total dataset into 90% training dataset and 10% evaluation dataset, our model achieves f1-score of 99.13% for the Ling-Spam dataset, and the confusion matrix for the evaluation dataset was as in figure7. Our spam detector model classifies correctly 378 ham emails as ham out of 379 input emails, while for spam class, our spam detector classifies correctly 81 emails as spams out of 84 spam emails.

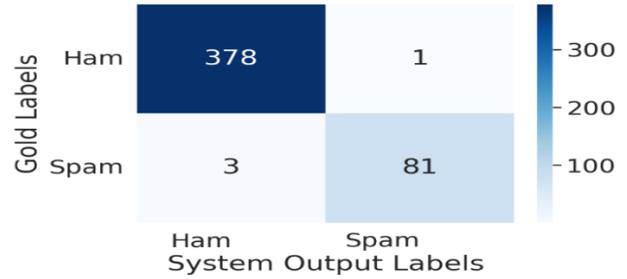

Figure 7: Confusion Matrix for Ling-Spam Corpus

### 6.1. SMS SPAM COLLECTION V.1 CORPUS RESULTS

We divide the corpus to 90% as a train dataset and the remaining 10% as evaluation dataset, and the resulting f1 score for the evaluation dataset was 99.28%, figure 8 shows the confusion matrix for evaluation dataset, and from the confusion matrix we can see that the system classifies 479 SMS as ham out of 481 ham SMS and classify 75 SMS as spam out of 77 SMS.

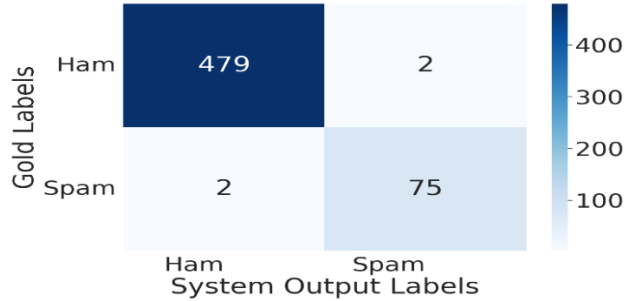

Figure 8: Confusion matrix for SMS Spam Collection Corpus

### 7. CONCLUSION

We have seen that spam is a serious issue in our day-to-day electronic communication over networks, that issue is not only an annoying issue to users, but it has a financial impact on business and individual users since emails or SMS can contain malicious malwares and phishing content that may lead to users' identity theft.

we have implemented an effective high performance spam detector that can detect spam emails or spam SMSs, we have trained our model on Enron corpus, SpamAssassin corpus, Ling-Spam corpus and SMS spam collection corpus, our spam detector performance was 98.62%, 97.83%, 99.13% and 99.28% respectively, the high performance of our model is a result of using BERT pre-trained model that enables our spam detector which allows our spam detector to better understand the message context and therefore better classification to spam or ham, that's why BERT outperform other machine learning algorithms in many tasks.

Using different corpuses to train and evaluate our model can allow any types of businesses to build their own spam detectors by simply collect a corpus that suits their business and train the model using it, and that solution would be a very suitable for startups.